\documentclass[fleqn,10pt]{wlscirep}
\usepackage[T1]{fontenc}
\usepackage{subfigure}
\usepackage{amsmath}
\usepackage{upgreek}
\usepackage{doi}
\DeclareCaptionLabelSeparator{vline}{ | }
\usepackage[singlelinecheck=false,figurename=Fig.,labelsep=vline,labelformat=simple]{caption}

\usepackage{lineno}

\usepackage{lipsum}

\usepackage{titlesec}

\newif\ifLNturnsON

\title{Exotic Spin-dependent Energy-level Shift Noise Induced by Thermal Motion}

\author[1]{Wei Xiao}
\author[1]{Xiyu Liu}
\author[1,*]{Teng Wu}
\author[1]{Xiang Peng}
\author[1,*]{Hong Guo}
\affil[1]{State Key Laboratory of Advanced Optical Communication Systems and Networks, School of Electronics, and Center for Quantum Information Technology, Peking University, Beijing 100871, China}

\affil[*]{corresponding author(s): Teng Wu (wuteng@pku.edu.cn); Hong Guo (hongguo@pku.edu.cn)}

\begin{document}
\begin{abstract}
Searching for exotic spin-dependent interactions that beyond the standard model has been of interest for past decades and is crucial for unraveling the mysteries of the universe. Previous laboratory searches primarily focus on searching for either static or modulated energy-level shifts caused by exotic spin-dependent interactions. 
Here, we introduce a theoretical model based on thermal motion of particles, providing another efficient way to search for exotic spin-dependent interactions. 
The theoretical model indicates that as the exotic spin-dependent interactions are related with the relative displacements and velocities of atoms,  atoms undergoing thermal motion would experience a fluctuating energy-level shift induced by the exotic interactions. Moreover, the resulting exotic energy-level shift noise could be sensed by high-sensitivity instruments. 
By using the model and taking the high-sensitivity atomic magnetometer as an example, we set the most stringent laboratory experiment constraints on eight different kinds of exotic spin- and velocity-dependent interactions, with five of which at the force range below 1 cm have not been covered previously. Furthermore, this theoretical model can be easily applied in other fields of quantum sensing, such as atomic clocks, atom interferometers and NV-diamond sensors, to further improve the laboratory constraints on exotic spin-dependent interactions.
\end{abstract}
\flushbottom
\maketitle

\thispagestyle{empty}

\noindent The undiscovered exotic spin-dependent interactions that beyond the standard model of particle physics have been of interest in past decades. The exotic interactions could potentially explain a number of unresolved significant problems in fundamental physics, such as the nature of dark matter \cite{Covi2001Axions,Bertone2005Particle,Safronova2018searchfor,Bertone2018History}, dark energy \cite{Friedland2003Domain,Flambaum2009Scalar}, the charge-parity violation in the strong interaction \cite{Peccei1977CP,Pierre2021Invisible}, and the hierarchy problem  \cite{Graham2015Cosmological}.

Since exotic spin-dependent interactions have been parameterized and extended in Ref. \citen{Dobrescu2006Spin}, introducing 16 kinds of interactions between ordinary particles: 15 of them are spin-dependent and 10 of them are velocity-dependent, many efforts have been made to uncover the exotic spin-dependent interactions.
Furthermore, given that these exotic interactions between particles can induce shifts in the energy levels of atoms, which could be detected by high-sensitivity instruments, various approaches have been made to search for the evidence of these exotic interactions \cite{Safronova2018searchfor,ficek2019constraining}, such as atomic magnetometers \cite{Kim2018Experimental,Feng2022Search,Wei2023femtotesla}, torsion
balances \cite{Ding2020Constraints},  nitrogen-vacancy (NV) centers in diamond \cite{Rong2018Constraints,Jiao2021Experimental}, magnetic microscopes \cite{Ren2021Search}, and other high-sensitivity technologies \cite{serebrov2010search,Guigue2015Constraining,Ficek2017Constraints,Ren2021Search}.


Previously, most of these laboratory search experiments focused on detecting static or modulated exotic energy-level shifts generated by either an unpolarized test mass, such as a ${\rm{Bi_4Ge_3O_{12}}}$ crystal \cite{kim2019experimental,Su2021Search} and the earth \cite{Search1992Venema,Kimball2013A,Zhang2023Search}, or a polarized test mass, such as SmCo5 permanent magnets \cite{Terrano2015Short,Wei2018New,Almasi2020New}.  These methods that based on energy-level shifts are definitely effective and have set increasingly stringent constraints on exotic spin-dependent interactions. However, in many scenarios, the energy-level shifts induced by other particles  tend to average out to zero, while the residual energy-level shift noise still exists and can also be used as a powerful tool to search for the exotic interactions.

Although, noise is usually unwelcome and its importance is often overlooked, it can actually provide valuable insights into the system itself.  Researchers can utilize the noise for detection purposes, such as exploring the spin dynamics through the spin-noise spectroscopy  \cite{crooker2004spectroscopy,Kolkowitz2015Probing,Sinitsyn2016The},  detecting single proteins using stochastic nuclear magnetic resonance signals \cite{Lovchinsky2016Nuclear,Pengfei2019Nanoscale}, and determining the temperature based on the inherent Johnson noise found across any resistor \cite{White1996status}.

Given that molecules, atoms and subatomic particles are almost always in a constant state of motion at temperatures exceeding absolute zero, and considering that the exotic interactions are predicted to be widely present among ordinary particles,  it is reasonable  to infer that particles in thermal motion would result in exotic energy-level shift noise within these particles. For high-performance sensors relying on atomic ensembles, the potential exists for the detection of such exotic energy-level shift noise by these sensors.
We take the following 10 specific spin- and velocity-dependent interactions as examples, 

\begin{equation}\label{eq:V45}
V_{4+5}=-f_{4+5} \frac{\hbar^{2}}{8 \pi m_{p} c}\left[\hat{\sigma}\cdot(\vec{v} \times \hat{r})\right]\left(\frac{1}{\lambda r}+\frac{1}{r^{2}}\right) e^{-r / \lambda},
\end{equation} 

\begin{equation}\label{eq:V1213}
V_{12+13}=f_{12+13} \frac{\hbar}{8 \pi}\left(\hat{\sigma}\cdot\vec{v} \right)\left(\frac{1}{ r}\right) e^{-r / \lambda},
\end{equation} 

\vspace{-0.3cm}
\begin{equation}\label{eq:V67}
V_{6+7}=-f_{6+7} \frac{\hbar^2}{4 \pi m_p c}\left[\left(\hat{\sigma}_1\cdot\vec{v} \right)\left(\hat{\sigma}_2\cdot\hat{r} \right)\right]\left(\frac{1}{ \lambda r}+\frac{1}{r^2}\right) e^{-r / \lambda},
\end{equation} 

\vspace{-0.3cm}
\begin{equation}\label{eq:V8}
V_{8}=f_{8} \frac{\hbar}{4 \pi c}\left(\hat{\sigma}_1\cdot\vec{v} \right)\left(\hat{\sigma}_2\cdot\vec{v}\, \right)\left(\frac{1}{ r}\right) e^{-r / \lambda},
\end{equation}

\vspace{-0.3cm}
\begin{equation}\label{eq:V14}
V_{14}=f_{14} \frac{\hbar}{4 \pi}\left[\left(\hat{\sigma}_1\times\hat{\sigma}_2\right)\cdot\vec{v}\,\right] \left(\frac{1}{ r}\right) e^{-r / \lambda},
\end{equation} 

\vspace{-0.3cm}
\begin{align}\label{eq:V15}
V_{15}=-f_{15} \frac{\hbar^{3}}{8 \pi m_{p}^2 c^2}\times\left\{\left[\hat{\sigma}_1\cdot(\vec{v} \times \hat{r})\right](\hat{\sigma}_2\cdot\hat{r})+(\hat{\sigma}_1\cdot\hat{r})\left[\hat{\sigma}_2\cdot(\vec{v} \times \hat{r})\right]\right\}\times\left(\frac{1}{\lambda^2 r}+\frac{3}{\lambda r^{2}}+\frac{3}{r^{3}}\right) e^{-r / \lambda},
\end{align}

\vspace{-0.3cm}
\begin{align}\label{eq:V16}
V_{16}=&-f_{16} \frac{\hbar^{2}}{8 \pi m_{p} c^2}\times\left\{\left[\hat{\sigma}_1\cdot(\vec{v} \times \hat{r})\right](\hat{\sigma}_2\cdot\vec{v})+(\hat{\sigma}_1\cdot\vec{v})\left[\hat{\sigma}_2\cdot(\vec{v} \times \hat{r})\right]\right\}\times\left(\frac{1}{\lambda r}+\frac{1}{ r^{2}}\right) e^{-r / \lambda},
\end{align}
where $f_i$ is a dimensionless coupling strength for the interaction $V_i$, $\hbar$ is reduced Planck’s constant, $m_p$ is the mass of the polarized particle, $c$ is the speed of light in vacuum, $\hat{\sigma}$ is the spin vector of the polarized particle, $\vec{v}$ is the relative velocity between the particles, $\hat{r}=\vec{r}/r$ is a unit vector of their relative displacement, and $\lambda=\hbar/m_ac$ is the force range, with $m_a$ being the axion mass.

\begin{figure}[bt!]
\centering
\includegraphics[width=0.94\linewidth]{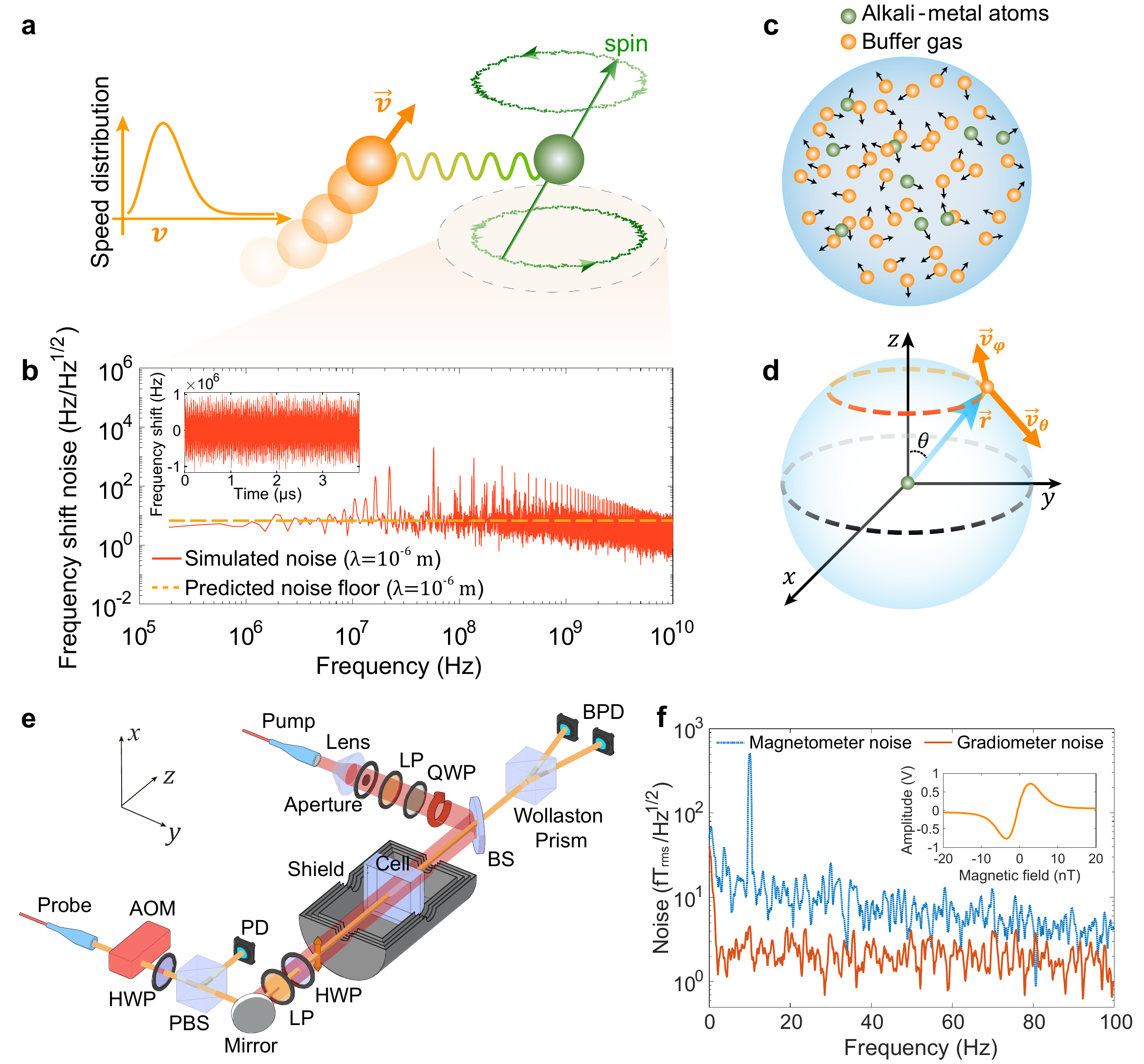}	
\!\!\subfigure{\label{fig:fig1A}}\!
\!\!\subfigure{\label{fig:fig1B}}\!
\!\!\subfigure{\label{fig:fig1C}}\!
\!\!\subfigure{\label{fig:fig1D}}\!
\!\subfigure{\label{fig:fig1E}}\!
\!\subfigure{\label{fig:fig1F}}\!
		\caption{{\bf{Schematic of the exotic energy-level shift noise induced by thermal motion of particles.}}
		\label{fig:fig1}
		{\bf{a}}, The velocity fluctuations of particles could generate exotic frequency shift noise due to the presence of spin- and velocity-dependent interactions, leading to random tilting of the spin of polarized atoms.
		{\bf{b}}, The simulated frequency shift noise induced by the interaction $V_{4+5}$ for Rb atoms confined in a $1\times1\times1$ ${\rm{\upmu m}}^3$ vapor cell filled with 1000 Pa buffer gas. To calculate the frequency noise detected by an atomic magnetometer, we assume a force range of $\lambda=$ 1 ${\rm{\upmu m}}$ and a coupling strength of $f_{4+5}=1$ for simplification (see Supplemental Materials for details). The simulated (solid line) and theoretical (dashed line) amplitude spectral density of the frequency noise induced by the interaction $V_{4+5}$. The predicted noise floor is consistent  with the simulated noise spectral density. The noise power is evenly distributed across the whole frequency range,  indicating that it follows a white noise characteristic.
		{\bf{c}}, An atomic vapor cell filled with alkali-metal atoms and buffer gas.
		{\bf{d}}, Schematic diagram of the exotic interaction $V_{4+5}$ between particles.
	{\bf{e}}, Schematic of the atomic magnetometers. A ring-shaped polarized light propagating along $z$ axis is performed as the pump light to polarize the atoms confined in an atomic vapor cell. 
	BS, beam splitter; LP, linear polarized; QWP, quarter-wave plate; PD, photodiode. For clarity, only one set of the atomic magnetometers is displayed in the figure.
		{\bf{f}}, The noise spectral density of atomic magnetometers developed in Ref. \citen{Wei2023femtotesla}. The inset shows the response of the magnetometer to the magnetic fields. Due to the magnetic field noise from the magnetic shield, the magnetometer shows a noise floor of 15 fT$_{\rm{rms}}$/Hz$^{1/2}$ (blue line). After suppressing the common-mode magnetic field noise in a gradiometer configuration, the gradiometer sensitivity reaches 2 fT$_{\rm{rms}}$/Hz$^{1/2}$ (red line).}
		
\end{figure}

The inherent random thermal motion of particles leads to fluctuations in their relative displacements and velocities between particles, consequently causing variations in the interaction strengths as indicated in Eqs. \eqref{eq:V45}$\sim$\eqref{eq:V16}.
As illustrated in Fig. \ref{fig:fig1A}, if the exotic interactions exist, these random variations could act as a source of noise for the exotic energy-level shift. For quantum sensors based on spins, these exotic energy-level shifts could randomly tile the atomic spins, and thus the resulting energy-level shifts or frequency shifts are detectable by high-sensitivity instruments such as atomic magnetometers \cite{kominis2003subfemtotesla}, atomic clocks  \cite{vanier2005atomic,Ludlow2015Optical}, superconducting quantum interference devices (SQUIDs) \cite{Fagaly2006Superconducting}, and NV centers in diamond \cite{Doherty2013NV,Barry2020Sensitivity}. 
As a result, these exotic interactions could set a fundamental noise limit for these quantum sensors, and in turn, the performance of these sensors can provide constraints on the characteristics of these exotic spin-dependent interactions.

In this paper, we propose a theoretical model based on thermal motion of particles, which can offer an alternative and efficient way to search for exotic spin-dependent  interactions. We exploit the atomic magnetometer as a high-sensitivity sensor to demonstrate the method for constraining the exotic spin-dependent  interactions based on the thermal motion of particles. 
As one of the most sensitive sensors for measuring magnetic fields \cite{kominis2003subfemtotesla}, the atomic magnetometer consists of a glass cell containing a vapor of alkali-metal atoms and laser lights for optically pumping or probing the atoms. 
For an ensemble of atoms in the thermodynamic equilibrium state, the velocity of atoms follows the Maxwell–Boltzmann velocity distribution \cite{auzinsh2010optically}. By utilizing this distribution, we have derived theoretical equations that describe the energy-level shift noise resulting from these exotic interactions shown in Eqs. \eqref{eq:V45}$\sim$\eqref{eq:V16}. Based on the theoretical model, we set the most stringent laboratory constraints on the coupling strength for these exotic interactions, which are several orders of magnitude tighter than previous constraints.

\subsection*{Thermal-motion-induced exotic frequency noise}
As shown in Fig. \ref{fig:fig1C}, we consider an atomic vapor cell filled with alkali-metal atoms and buffer gas. To estimate the exotic spin-dependent interactions between these particles, we concentrate on a certain specific alkali-metal atom and derive the fluctuations of interaction strengths in the reference frame of the alkali-metal atom. For atomic spins, the atomic energy-level shift results in transition frequency shifts.
As an illustrative example, we examine the exotic frequency shift noise of $V_{4+5}$ induced by the thermal motion.
The total exotic interactions between the alkali-metal atom and other particles can be described as
\begin{subequations}
\begin{eqnarray}
V_{4+5}^{\rm{tot}}\!&=&\!\hbar\hat{\sigma}\cdot{\vec{\omega}}_{4+5},\\
{\vec{\omega}}_{4+5}\!&=&\!-\frac{f_{4+5}\hbar}{8 \pi m_{p} c}\sum_i^N (\vec{v}_i \times \hat{r}_i)\left(\frac{1}{\lambda r_i}+\frac{1}{r_i^{2}}\right) e^{-r_i / \lambda},\quad\quad\label{eq:w45_1}
\end{eqnarray}
\end{subequations}
where $\vec{v}_i$ and $\vec{r}_i$ are the relative velocity and displacement between the $i$th  particle and the alkali-metal atom, respectively. ${\vec{\omega}}_{4+5}$ is the total frequency shift  induced by the exotic interaction between the atom and other $N$ particles.

The relative velocity $\vec{v}_i$ can be expanded as $\vec{v}_i=v_r\hat{r}+v_\theta\hat{\theta}+v_\varphi\hat{\varphi}$ in spherical polar coordinates, where $\theta$ and $\varphi$ are the polar angle and azimuthal angle, respectively, as shown in Fig. \ref{fig:fig1D}.
For simplification but without loss of generality, we consider the $z$-axis component of the frequency shift ${\omega}_{4+5}^{(z)}$.
In this way, the total frequency shifts can be derived by summing all the interactions between the atom and other $N$ particles.
When the atomic vapor cell is uniformly heated with a negligible temperature gradient around the cell, particles are uniformly distributed throughout the entire space and follow the Maxwell–Boltzmann velocity distribution. The total frequency shift produced by particles in the range of radius $\delta R<r<R$ is
\begin{equation}
{\omega}_{4+5}^{(z)}=-f_{4+5}\frac{\hbar}{8 \pi m_{p} c}\!\int_{-\infty}^{\infty}\int_{\delta R}^R\int_0^{\pi}\int_0^{2\pi}\!nr^2\sin\theta{\rm{d}}\varphi {\rm{d}}\theta{\rm{d}}r\cdot{v}_\varphi\sin\theta \left(\frac{1}{\lambda r}+\frac{1}{r^{2}}\right) e^{-r/ \lambda}\rho({v}_\varphi )\mathrm{d}{v}_\varphi ,\label{eq:w45_3}
\end{equation}
where $n$ is the particle number density and $\rho(v_\varphi)$ is the velocity distribution function.

According to Eq. \eqref{eq:w45_3}, it can be derived that the average value of the total frequency shift is zero,
\begin{equation}
E\left[{\omega}_{4+5}^{(z)}\right]=0,
\end{equation}
because the integrand about the velocity is an odd function and the integral interval is symmetric about 0.
However, the fluctuation of the total frequency shift still exists and its standard deviation can be expressed as 
\begin{subequations}
\begin{eqnarray}
{\sigma}\left({\omega}_{4+5}^{(z)}\right)\!\!\!&=&\!\!\!\sqrt{E\left[({\omega}_{4+5}^{(z)})^2\right]-E\left[{\omega}_{4+5}^{(z)}\right]^2}=\left(\frac{f_{4+5}\hbar}{8 \pi m_{p} c}\right)\sqrt{\frac{8\pi nk_{\rm{B}}T}{3\mu}\left[{\eta}(\delta R)-{\eta}(R)\right]},\quad\ \label{eq:V45STD}\\
\eta(r)\!\!\!&=&\!\!\!\left(\frac{1}{2\lambda}+\frac{1}{r}\right)e^{-{2 r}/{\lambda }},
\end{eqnarray}
\end{subequations}
where $k_B$ is the  Boltzmann constant, $T$ is the cell temperature, and $\mu$ is the reduced mass of the particle and the atom. Detailed derivations are presented in Supplemental Materials.

With the standard deviation of the noise, the bandwidth of the noise should be obtained as well, otherwise, we can not extract the strength of the frequency shift noise. Since the power spectral density function and the autocorrelation function $R(\tau)$ of a signal form a Fourier transform pair, we can estimate the bandwidth of the noise by calculating the autocorrelation function 
\begin{equation}
R(\tau)=E\left[{\omega}_{4+5}^{(z)}(0){\omega}_{4+5}^{(z)}(\tau)\right],\label{eq:BW}
\end{equation}
where the fluctuation noise is assumed to be a stationary process. The derivation of the bandwidth is much more complex and we can not obtain an analytical expression of it. So, the bandwidth is extracted by numerical integrations (see Supplemental Materials for further details).
Since the random thermal motion of particles is the source of common thermal noise (or Johnson–Nyquist noise) \cite{Johnson1928Thermal,Nyquist1928Thermal}, which is usually considered to be approximately white, it is reasonable to assume that the thermal-motion-induced frequency shift noise could also be white, meaning that the amplitude spectral density is nearly constant within the frequency bandwidth. In this case, we finally obtain the spectral density of the exotic frequency shift noise, which can be denoted as
\begin{equation}\label{eq:w45_ASD}
\begin{split}
\delta{{\omega}_{4+5}^{(z)}}&=\frac{{\sigma}\left({\omega}_{4+5}^{(z)}\right)}{\sqrt{\rm{BW}}}.
\end{split}
\end{equation}

As $z$ direction is arbitrarily specified, the frequency shift noise exists isotropically in all directions. 
The results suggest that for the atoms undergoing thermal motion, the velocity-dependent interaction would randomly tile the atomic spins in any directions, as shown in Fig. \ref{fig:fig1A}. Since the thermal motion is inevitable in almost all of the atom-based devices, the thermal-motion induced frequency shift noise can be widely present in all of these devices. In highly sensitive sensing devices, the exotic frequency noise can drown out weak signals, and can be a limiting factor on the performance of the measuring instruments. As the exotic frequency shift noise increases with temperature, its strength can be reduced by cooling down the instruments. 

Additionally, not only the exotic interaction $V_{4+5}$ induce frequency shift noise, all the exotic interactions that depend on the velocity cause similar types of frequency shifts or frequency shift noises. 
Specifically, the exotic interaction $V_{4+5}$, $V_{12+13}$, $V_{6+7}$, $V_{14}$, $V_{15}$, and $V_{16}$ would only cause frequency shift noise with zero-mean shift, while the exotic interaction $V_{8}$ would cause a non-zero constant frequency shift along the direction of the spin polarization. 
Because the interaction $V_8$, as shown in Eq. \eqref{eq:V14}, is proportional to the square of the velocity. This indicates that the integrand with respect to the velocity is an even function when calculating the average total frequency shift $\vec{\omega}_{8}$, similar to the way used in Eq. \eqref{eq:w45_3}.
The theoretical expressions for these frequency shifts have been derived, and detailed derivations can be found in Supplemental Materials.

\subsection*{High-sensitivity atomic magnetometer}
Having previously demonstrated that the different exotic spin- and velocity-dependent interactions would cause different types of frequency shift noises, the detection of such exotic frequency shift noise can be accomplished using various quantum sensors. 
Here, we exploit the atomic magnetometer, which is one of the most sensitive instruments for magnetic field measurements, as a high-sensitivity sensor to search for the exotic frequency shift noise. Since the atomic magnetometer is operated to detect the magnetic field by measuring the energy-level splitting induced by the Zeeman effect, it exhibits sensitivity not only to magnetic fields but also to frequency shift noise. So, we consider an atomic magnetometer that we developed previously in Ref. \citen{Wei2023femtotesla}, as shown in Fig. \ref{fig:fig1E}.
The atoms confined in the vapor cell are optically pumped with a circularly-polarized and ring-shaped light. To measure the effective magnetic field along $y$ axis, we adopt a transverse parametric resonance  scheme to extract the magnetic field information from the atoms. The dynamic evolution of the spin polarization ${\bf{P}}$ can be described with the Bloch equation,
\begin{equation}\label{eq:BlochEqu}
\frac{{\rm{d}} }{{\rm{d}} t}{\bf{P}}=\gamma {\bf{P}} \times {\bf{B}}+R_{\rm{OP}}({\bf{s}}-{\bf{P}})-\Gamma_2{\bf{P}},
\end{equation}
where $\gamma$ is the gyromagnetic ratio, {\bf{B}} is the magnetic field, $R_{\rm{OP}}$ is the optical pumping rate that is proportional to the light intensity, ${\bf{s}}$ is the optical pumping vector that defines the direction and the polarization of the light,  and $\Gamma_2$ is the spin relaxation rate. 

Under a transverse modulated field $B_1\cos\omega_{\rm{rf}} t$ along $y$ axis, the spin response along $z$ axis can be derived with Eq. \eqref{eq:BlochEqu}, which is
\begin{equation}\label{eq:RFmodulation}
P_z(\omega_z)=2P_0J_0(\eta)J_1(\eta)\frac{\Gamma \gamma B_{0y}}{\Gamma^2+\gamma^2B_{0y}^2}\sin\omega_{\rm{rf}} t,
\end{equation}
where  $P_0$ is the equilibrium spin polarization in the absence of magnetic fields, $J_n$ is the Bessel function of order $n$, $\eta=\gamma B_1/\omega_{\rm{rf}}$ is the modulation index, and $B_{0{y}}$ is the $y$-axis magnetic field to be measured. 
Since the optical rotation measured with the probe light is proportional to the component of the  spin polarization in the probe propagation direction, the measured optical rotation is a form of dispersive Lorentzian curve centered around $B_{0y}=0$, as indicated in Eq.~\eqref{eq:RFmodulation}. The measured response to the magnetic fields is depicted in the inset of Fig. \ref{fig:fig1F}.

Due to the thermal magnetic field noise generated by the innermost layer of the magnetic shield ($\mu$-metal), a two-channel OPM is developed to suppress the common mode noise from the background. Based on the gradiometer configuration, the atomic magnetometer achieves a noise floor of  $\sim$2 fT$_{\rm{rms}}$/Hz$^{1/2}$, as shown in Fig. \ref{fig:fig1F}. Detailed experimental setups can be found in Ref. \citen{Wei2023femtotesla}. Such a magnetic-field sensitivity indicates a frequency noise floor of $\sim$10 $\upmu {\rm{Hz}}_{\rm{rms}}$/Hz$^{1/2}$, which can be used to establish an upper bound to the exotic frequency noise.

\subsection*{Constraints on exotic interactions}
As shown in Eq. \eqref{eq:w45_ASD}, an atom in a vapor cell undergoing thermal motion would sense the exotic frequency shift noise induced by other particles. However, for an atomic magnetometer based on light-atom interactions, there usually involves many atoms rather than a single atom. 
Therefore, the frequency noise detected by an atomic magnetometer is the average of the noise of all atoms interacting with light. 
Since the thermal-motion-induced frequency noise is related to the particle velocity, as illustrated  in Eqs. \eqref{eq:V45}$\sim$\eqref{eq:V16}, and the velocity of each particle is independent, we naturally assume that the noise sensed by each alkali-metal atom is independent and identical. Under these conditions, the spectral density of the fluctuation noise detected by an atomic magnetometer can be denoted as
\begin{equation}\label{eq:w45_AveASD}
\overline{\delta{\omega}}_{4+5}^{(z)}=\frac{{\delta{\omega}}_{4+5}^{(z)}}{\sqrt{N_{\rm{int}}}}=\frac{{\sigma}\left({\omega}_{4+5}^{(z)}\right)}{\sqrt{N_{\rm{int}}\cdot\rm{BW}}},
\end{equation}
where $N_{\rm{int}}$ is the number of atoms interacting with light.

Equation \eqref{eq:w45_AveASD} indicates that the random thermal motion of particles induces frequency shift noise that depends on the characteristics of the exotic interaction, such as the coupling strength and the force range. Therefore, if  the exotic interaction exists, these frequency shift noises would be sensed by the atomic magnetometer and thus would introduce another fundamental noise limit distinct from photon shot noise or spin-projection noise\cite{Sorensen1998Quantum} on atomic magnetometers. 

When the coupling strength of exotic spin-dependent interactions is sufficiently large, it is possible to observe an unidentified noise source that affects the sensitivity of atomic magnetometers differently from photon shot noise and spin-projection noise. 
Moreover, this unknown noise source exhibits an increase in noise floor with buffer gas pressure and cell size, as shown in Eq. \eqref{eq:V45STD}. Under such circumstances, it is reasonable to suspect that the unknown noise arises from the exotic spin-dependent interactions and we can further design a more targeted experiment to verify the origins of the noise.

\begin{figure*}[hbtp]
\centering
\subfigure{\label{fig:fig2A}}
\subfigure{\label{fig:fig2B}}
\includegraphics[width=0.97\linewidth]{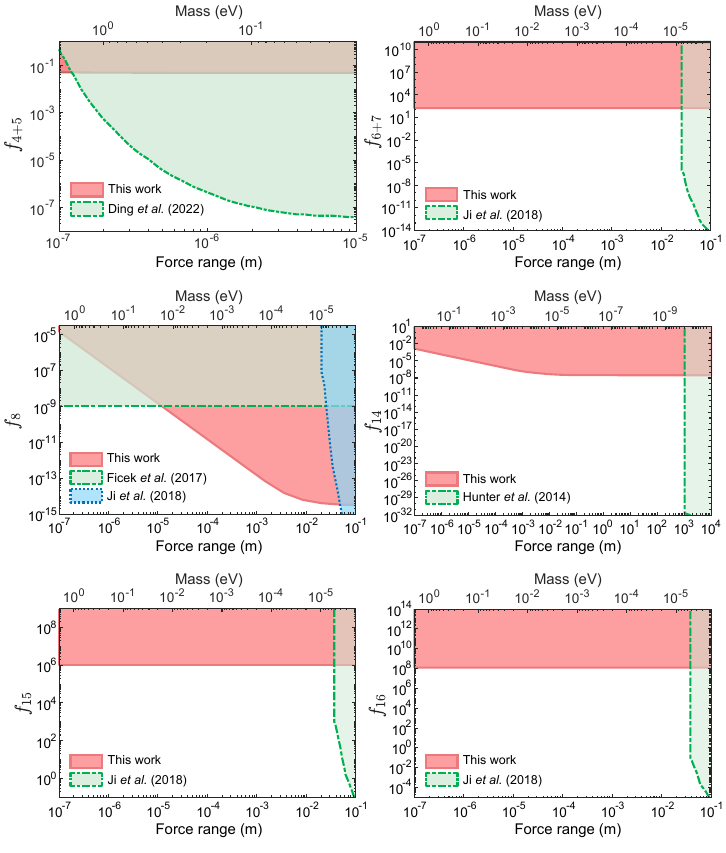}	
		\caption{{\bf{Constraints on the coupling strengths of exotic interactions as a function of the force range}}.
		The references for different terms of the velocity-dependent interactions are: $V_{4+5}$ from Refs. \citen{Ding2020Constraints},  $V_{6+7}$ from Ref. \citen{Wei2018New}, $V_8$ from Ref. \citen{Ficek2017Constraints,Wei2018New}, $V_{14}$ from Ref. \citen{Hunter2014Using}, $V_{15}$ and $V_{16}$ from Ref. \citen{Wei2018New}.}
		\label{fig:constraints}
\end{figure*}

It should be noted that atoms in thermal motion may be influenced by different types of exotic spin-dependent interactions. However, regardless of what kind of exotic interactions exist, all of these different frequency shift noises contribute to a total frequency shift noise. Since each frequency shift noise has distinct characteristics, as illustrated in Eqs. \eqref{eq:V45}$\sim$\eqref{eq:V16}, it is assumed that these frequency shift noises are independent. Consequently, the total frequency shift noise is greater than any individual noise source. In this case, the noise floor achieved in atomic magnetometers can in turn establishes an upper bound for the total noise strength, thus also setting an upper bound to each of these exotic frequency shift noises.

To be reasonable, we only consider particles in a spherical space with a radius of $R=$ 1 mm (a typical value of the light diameter), and exclude the space with a radius smaller than  $\delta R=10$ nm (approximately 40 times the atomic radius) since the exotic interaction formulas need to be modified at the atomic scale \cite{Fadeev2019Revisiting}. Furthermore, to ensure there are enough particles within the force range and make the velocity fluctuations meet the requirements of statistical analysis, i.e., $n\lambda^3\gg1$, the force range discussed in this thermal-motion model is limited to $\lambda>10^{-7}$ m.

Given the experiment parameters and theoretical model, we set constraints on the coupling strengths of interactions $V_{4+5}$, $V_{12+13}$, $V_{6+7}$, $V_{8}$, $V_{14}$, $V_{15}$, and $V_{16}$. For exotic interactions $V_{4+5}$ and $V_{12+13}$ between polarized particles and unpolarized particles, the electron spins of Rb atoms are the polarized particles and the buffer gas N$_2$ serves as the unpolarized mass source.
For other exotic interactions between two polarized particles, the electron spins of Rb atoms themselves serve as the polarized spin source.

As shown in Fig. \ref{fig:constraints}, our works set the most stringent constraint on the coupling strength of the spin- and velocity-dependent interaction $V_{4+5}$ between Rb electron spins and unpolarized nucleons for the force range around $10^{-7}$ m, and improve the constraint on the interaction $V_8$ between Rb electron spins by several orders of magnitude for the force range between  $1\times10^{-5}$ m and $5\times10^{-2}$ m. 
Moreover, we also establish the first limit on the velocity-dependent interactions  $V_{6+7}$, $V_{14}$, $V_{15}$, and $V_{16}$ between electron spins for the force range between $10^{-7}$ m and $10^{-2}$ m. For the exotic interaction  $V_{12+13}$, the previous constraints are more stringent than the constraint set by our work and the detailed constraints can be found in Supplemental Materials.

\section*{Discussions}
If we take a more sensitive atomic magnetometer or comagnetometer as an example, such as the sub-femotesla atomic magnetometers built previously   \cite{allred2002high,kominis2003subfemtotesla,sheng2013subfemtotesla}, 
the constraints on these interactions might be further improved by about one order of magnitude.
Since the exotic frequency shift noise covers a wide frequency range, as shown in Fig. \ref{fig:fig1B}, the radio-frequency (rf) atomic magnetometers \cite{Lee2006NQR} are also applicable to search for such exotic interactions.
As for the shot-noise limited atomic magnetometer \cite{allred2002high}, for example, the estimated shot-noise limit of the spin-exchange relaxation free (SERF) magnetometer is $\sim$2 aT/Hz$^{1/2}$, and three orders of magnitude extra improvement on the constraints shown in Fig. \ref{fig:constraints} can be expected.

In conclusion,  we have proposed a theoretical model describing the frequency shift noise induced by exotic spin-dependent interactions due to the thermal motion of particles. This exotic frequency shift noise can be sensed by high-sensitivity instruments and serves as a fundamental noise limit on these instruments. 
Taking a high-sensitivity atomic magnetometer that we previously developed in Ref. \citen{Wei2023femtotesla} as an example, we have calculated the expected frequency shift noise sensed by the atomic magnetometer. Consequently, we in turn set a limit on the coupling strength of these exotic interactions according to the noise floor we achieved with the atomic magnetometer. 

Since our theoretical model is based on the thermal motion of microscopic particles, it naturally allows for the exploration of exotic spin-dependent interactions on a small scale that has not been investigated yet.
Through our investigations, we have significantly  improved the constraints on exotic interactions $V_{4+5}$ and $V_{8}$ by several orders of magnitude. Additionally, we have established the first limits on the exotic interactions $V_{6+7}$, $V_{14}$, $V_{15}$, and $V_{16}$ for the force range between $10^{-7}$ m and $10^{-2}$ m. Notably, for interactions involving polarized and unpolarized particles, i.e., $V_{4+5}$ and $V_{12+13}$, except for the buffer gas filled in the atomic vapor cell can be served as the unpolarized mass source, we can even take the air on the earth or the cosmic dust as the mass source of thermally moving particles in principle, which would extend the force range shown in Fig. \ref{fig:constraints} and yield even tighter constraints on the coupling strengths over a broader scale.


Furthermore, we believe that the theoretical model proposed in this paper is not confined to atomic magnetometers and could be applied to various other fields of quantum sensing. For example, the atomic clock that is based on the precise detection of the energy difference between two levels might also sense the thermal-motion-induced frequency shift noise. Moreover, the vibration of crystal lattices might cause similar exotic frequency shift noise to be sensed by NV-based quantum sensors \cite{Doherty2013NV,Barry2020Sensitivity}. 
We believe that the analysis presented in this paper can also be adapted to set boundaries on exotic interactions in many other fields, broadening the potential applications and implications of our research.
This research advances our understanding of these exotic spin-dependent interactions and remains us the noise can also be a powerful tool in exploring and constraining exotic physics beyond the standard model.
\bibliography{reference}

\section*{Methods}
The high-sensitivity atomic magnetometer we developed is a SERF Rb magnetometer based on transverse parametric resonances. The atoms are optically pumped by a circularly-polarized and ring-shaped light, and the atomic spin polarization is detected with a linearly polarized light that is red detuned about 60 GHz away from the center of the Rb D1 line. To suppress the intensity noise of probe light, its power is stabilized with an acousto-optic modulator (AOM). The detailed experiment parameters are listed in Table \ref{tab:parameters}. 
\begin{table}[bt]
\renewcommand{\arraystretch}{1.25}
\caption{Experimental parameters used for the estimation of thermal-motion-induced exotic frequency shifts.}\label{tab:parameters}
\setlength{\tabcolsep}{6.5mm}{
\begin{tabular}{lc}
\hline \hline Parameter & Value \\
\hline
Magnetometer sensitivity & 2 fT$_{\rm{rms}}$/Hz$^{1/2}$\\
Species & $^{87}{\rm{Rb}}$ \\
Buffer gas & 600 Torr N$_2$\\
Cell temperature & 153$^\circ$C\\
Spin polarization & 0.5\\
Minimal radius $\delta R$ & 10 nm\\
Maximum radius $R$ & 1 mm\\
\hline
\hline
\end{tabular}}
\end{table}

To further verify the theoretical model and the derived theoretical expressions for these exotic frequency shift noise, we simulated the random thermal motion of particles and extracted the simulated frequency shift noises, as shown in Fig. \ref{fig:fig1B}. Due to limitations in computing resources, it is impractical to directly simulate all the atoms in a centimeter-sized atomic vapor cell ($\sim$$10^{19}$ atoms for 600 Torr buffer gas) that conforms to actual experimental parameters. 
Although, we have reduced the cell size to 1 $\upmu$m and the pressure of the buffer gas to 1000 Pa, the computation speed is still significantly slow due to the tiny time step and huge particle number ($\sim$170,000) in simulation. After several weeks of simulations, we simulated the atomic thermal motion for a total time length of approximately  3.8 $\upmu$s.
To calculate the frequency shift noise detected by the atomic magnetometer, we choose 100 Rb atoms as the sensing atoms that interacting with light.  
The standard deviation calculated by using Eq. \eqref{eq:V45STD} is found to be in agreement with the simulation results, as shown in Fig. \ref{fig:fig1B}. 
Furthermore, the noise bandwidth derived by Eq. \eqref{eq:BW} is also consistent with the simulation results. The simulated noise spectral density depicted in Fig. \ref{fig:fig1B} also supports the assumption of white noise and thus the predicted noise floor agrees with the simulated noise floor. For more details about the simulation, please refer to Supplemental Materials.

\section*{Data availability}
All data needed to evaluate the conclusions in the paper are present in the paper and/or the Supplementary Materials. Other data that support the findings of this study are available from the corresponding author upon reasonable request.

\section*{Code availability}
The code that supports the plots in this paper is available from the corresponding author
upon reasonable request.

\section*{Acknowledgements}
This work is supported by the National Natural Science Foundation of China (Grants No. 62071012, No. 61571018, No. 61531003, No. 91436210), the National Science Fund for Distinguished Young Scholars of China (Grant No. 61225003), and National Hi-Tech Research and Development (863) Program. T. W. acknowledges the support from the start-up funding for young researchers of Peking University.

\section*{Author contributions}
W.X. conceived the idea. H.G., X.P. and T.W. guided the project. W.X. and
X.Y. L. carried out the theoretical calculations. W. X. performed the numerical simulation. 
All authors discussed the results and commented on the manuscript.

\section*{Competing interests}
The authors declare no competing interests.

\end{document}